\documentclass[twocolumn]{ceurart}

\usepackage{listings}
\begin{document}
    
    \copyrightyear{2025}
    \copyrightclause{Copyright for this paper by its authors.
        Use permitted under Creative Commons License Attribution 4.0
        International (CC BY 4.0).}
    
    \conference{12th Joint Workshop on Interfaces and Human Decision Making for Recommender Systems (IntRS’25)}
    
    \title{Towards LLM-Enhanced Group Recommender Systems}
    
    \author[1]{Sebastian Lubos}[%
    orcid=0000-0002-5024-3786,
    email=sebastian.lubos@tugraz.at,
    url=https://ase.sai.tugraz.at/,
    ]
    \cormark[1]
    \address[1]{Graz University of Technology, Inffeldgasse 16b, Graz, 8010, Austria}
     \address[2]{Uniquare Software Development, Lannerweg 9, Krumpendorf, 9201, Austria}
      \address[3]{Innovation Service Network, Grabenstraße 231, Graz, 8045, Austria}
       \address[4]{Morgendigital,  Leopoldstraße 20, Innsbruck, 6020, Austria}

    \author[1]{Alexander Felfernig}[%
    orcid=0000-0003-0108-3146,
    email=alexander.felfernig@tugraz.at,
    url=https://ase.sai.tugraz.at/,
    ]

    \author[1]{Thi Ngoc Trang Tran}[%
    orcid=0000-0002-3550-8352,
    email=ttrang@ist.tugraz.at,
    url=https://ase.sai.tugraz.at/,
    ]
    
    \author[1]{Viet-Man Le}[%
    orcid=0000-0001-5778-975X,
    email=v.m.le@tugraz.at,
    url=https://ase.sai.tugraz.at/,
    ]

    \author[1]{Damian Garber}[%
    orcid=0009-0005-0993-0911,
    email=damian.garber@tugraz.at,
    url=https://ase.sai.tugraz.at/,
    ]

    \author[2]{Manuel Henrich}[%
    email=Manuel_Henrich@uniquare.com,
    url=https://www.uniquare.com/,
    ]

    \author[3]{Reinhard Willfort}[%
    email=reinhard.willfort@innovation.at,
    url=https://innovation.at,
    ]

     \author[4]{Jeremias Fuchs}[%
    email=jerry@morgendigital.com,
    url=https://www.morgendigital.com/,
    ]
    
    \cortext[1]{Corresponding author.}

\begin{abstract}
In contrast to single-user recommender systems, group recommender systems are designed to generate and explain recommendations for groups. This group-oriented setting introduces additional complexities, as several factors—absent in individual contexts—must be addressed. These include understanding group dynamics (e.g., social dependencies within the group), defining effective decision-making processes, ensuring that recommendations are suitable for all group members, and providing group-level explanations as well as explanations for individual users. In this paper, we analyze in which way large language models (LLMs) can support these aspects and help to increase the overall decision support quality and applicability of group recommender systems.
\end{abstract}


\maketitle

\section{Introduction}

Nowadays, recommender systems are a foundational component of many digital platforms, which help to personalize item suggestions \cite{JannachetalRecSys2010}. Traditionally, recommender systems analyze the preferences and behaviors of \emph{single users} to optimize item rankings. In contrast, \emph{group recommender systems} \cite{Jameson2007,Masthoff2011,felfernig2024group} support the decision making of multiple users by aggregating their individual preferences. This approach triggers various challenges such as the resolution of conflicting preferences, the assurance of fairness with regard to the determined recommendations, and the identification of appropriate decision (aggregation) strategies. While recommender systems for single users focus on personalization, group recommender systems must also take into account the aspects of negotiation (among users) and compromise with regard to the proposed items.

The application of group recommender systems is relevant in the context of decision tasks with a focus on \emph{collaborative decision-making} \cite{Jamesonetal2022}. For example, families or groups of friends commonly make joint decisions about holiday destinations, accommodations, and itineraries. In a similar fashion, restaurant selection is often performed by groups (e.g., restaurant selection for a Christmas party). The group decision support tool \textsc{Doodle}\footnote{https://doodle.com/} can be regarded as a very basic form of group recommender system where preferred time slots can be interpreted as implicit recommendation (based on majority voting) \cite{Reinecketeal2013Doodle}. Although entertainment environments such as Netflix\footnote{\url{https://netflix.com}} have already performed initial studies regarding the applicability of group recommender systems \cite{BerryetalNetflixGroupRecommender2010}, their productive platforms are still focused on individual users, i.e., not groups. Music recommendation provides another relevant application domain for group recommender systems, for example, in terms of shared listening environments for parties or gyms. However, popular platforms like Spotify\footnote{\url{https://www.spotify.com}} have not integrated group recommendation with the exception of basic collaborative playlists \cite{Kwaketalspotifygrouprecommendation2024}. In software engineering, group recommender systems are applied specifically in the context of requirements engineering scenarios where the preferences and background knowledge of stakeholders are an input for recommender systems that support the prioritization of software requirements   \cite{Ninausetal2014IntelliReq}. For an overview of applications of group recommender systems, we refer to Felfernig et al. \cite{felfernig2024group}.

Despite the existence of different promising application areas for group recommender systems, these systems are not widely applied and accepted. On the one hand, a reason behind is a \emph{lack of flexibility} in the support of group decision processes, for example, in taking over a more supportive role in contrast to the proposal of concrete recommendations \cite{jannach2025rethinkinggrouprecommendersystems}. On the other hand, group decisions are often based on \emph{sensitive or private information} that people do not want to share with others in an explicit fashion. Examples thereof are physical condition in holiday round trip planning, interpersonal likings in group formation or personal selection, or personal preferences regarding movies \cite{LuoChen2014PrivacyPreservingGroupRecommendation2014}. Furthermore, some group decision scenarios suffer from \emph{hidden agendas}, where individuals strategically withhold or distort their preferences to manipulate the group outcome \cite{TrangetalManipulationCounteractionInterfacesGroupRec2019}. Examples thereof can be found in professional environments, for example, strategy meetings, recruitment sessions, or funding decisions. In contrast, domains involving leisure and less-sensitive content, such as restaurants and meeting dates, show a higher acceptance of group decision support.

In addition to technical, privacy, and manipulation issues, \emph{cultural and societal values} could also play a role in the adoption of group recommender systems \cite{hofstede2001culture,ZhangetalIndividualismCulture2007}. In Western societies, we can observe a tendency towards individualism, which can translate to a strong focus on individual preferences. This can also counteract the acceptance of group recommender systems, which require, to some extent, a preparedness to compromise and openness. On the other hand, in more collectivist cultures, group harmony is prioritized higher, which could increase an interest in group recommender systems. As a consequence, not only technical performance but also sociocultural aspects have to be regarded as impact factors for the acceptance of group recommender systems.

Finally, the internal structure of the group (i.e., homogeneous vs. heterogeneous) has an impact on group recommender acceptance. In homogeneous groups such as close friends with shared tastes, basic preference aggregation could be an accepted recommendation mechanism. Heterogeneous groups such as large (often cross-cultural) software development teams might be confronted with diverse and often conflicting preferences. In such scenarios, simple aggregation functions reach their limits, and more sophisticated decision support is needed that considers aspects such as fairness, the need for negotiating decisions, and resolving conflicts. In this paper, we focus on the aspect of making the decision support provided by group recommender systems more flexible. In the line of Lin et al. \cite{Linetal2025} and Zhang et al. \cite{Zhangetal2024}, we analyze \emph{different ways of how large language models (LLMs) can support group decision scenarios}.

The remainder of this paper is organized as follows. Section 2 provides an overview of basic algorithms for group recommender systems and highlights how LLMs can enhance these algorithms. In Section 3, we discuss existing preference elicitation methods for group recommenders and how these methods can be improved on the basis of LLMs. Section 4 outlines how LLMs can support explanations in the context of group recommendation scenarios. Section 5 shows how synergy effects can be achieved by combining LLMs with insights from psychological theories of decision making. Section 6 summarizes open research challenges related to the integration of LLMs into group recommender systems. The paper is concluded in Section 7.

\section{Algorithms for Group Recommender Systems}

\emph{Group-based collaborative filtering} (GCF) extends basic collaborative filtering for single users by aggregating the preferences of single users into a group model.\footnote{For a detailed discussion of different related group recommendation approaches we refer to Felfernig et al. \cite{felfernig2024group,BorattoFelfernigGroupReccollaborativeRec2018,Masthoff2022}.} Examples of such aggregation strategies are the \emph{average} strategy (the mean of all user preferences), \emph{least misery} (the satisfaction of the least happy user), and most pleasure (the satisfaction of the happiest user). Consider a group of three users rating four movies as shown in Table~\ref{tab:gcf}. Using least misery, the group score for Movie A would be $\min(4, 5, 2) = 2$, while the average strategy would result in $\frac{4+5+2}{3} = 3.67$. Such aggregation functions, in many cases, oversimplify group dynamics and the history of the decision process. 

\begin{table}[h]
\centering
\caption{Example of group member (user) ratings for four movies.}
\label{tab:gcf}
\begin{tabular}{lcccc}
\toprule
\textbf{User} & \textbf{Movie A} & \textbf{Movie B} & \textbf{Movie C} & \textbf{Movie D} \\
\midrule
User 1 & 4 & 3 & 5 & 2 \\
User 2 & 5 & 4 & 4 & 1 \\
User 3 & 2 & 4 & 5 & 3 \\
\bottomrule
\end{tabular}
\end{table}

\emph{Content-based filtering for group recommendation} (CBFG) is based on the idea of constructing and aggregating user profiles \cite{Perezetalcontentbasedgrouprec2021}, for example, as vectors of preferred item features. If User A likes movies with the features ``romantic'' and ``comedy'' and User B prefers ``action'' and ``thriller'', a combined profile must balance these dimensions. Let the profiles of two users be $\vec{u}_A = [1, 0.8, 0, 0]$ and $\vec{u}_B = [0, 0, 1, 0.9]$ for the genres [romance, comedy, action, thriller]. A simple average profile is $[0.5, 0.4, 0.5, 0.45]$. 

\emph{Critiquing-based group recommendation} (CRITG) -- in contrast to GCF and CBFG -- supports user feedback cycles in terms of critiques \cite{McCarthyetalgrouprecCritiquing2006}. Examples of critiques in restaurant recommendation are ``less expensive'', ``nearer'', or ``higher food quality''. Such critiques are aggregated where conflict resolution can become an issue due to conflicting critiques provided by different users (e.g., ``high quality food'' vs. ``less expensive'' could lead to a situation where no recommendation can be identified) \cite{felfernig2024group}.

\textbf{LLM-related Potentials.} LLMs have the potential to significantly enhance algorithmic approaches for group recommender systems by interpreting user input in a more human-centered way. In the context of GCF, LLMs can move beyond numerical ratings by \emph{interpreting free-form feedback} such as “I loved the plot but hated the violence” and \emph{transforming it into preference vectors}. In CBFG, LLMs can \emph{extract item attributes} from unstructured sources with the goal of enriching user and item profiles. In the context of CRITG, LLMs can \emph{mediate conflicting critiques} using generative dialog strategies that \emph{suggest reformulations}. Most importantly, LLMs can be exploited to dynamically \emph{propose appropriate preference aggregation and decision strategies}.

\section{Preference Elicitation in Group Recommender Systems}

In the context of group recommender systems, the complexity of preference elicitation increases compared to single-user recommender systems due to the presence of different and possibly conflicting preferences. In group decision contexts, preferences can be elicited by collecting explicit ratings from users (i.e., group members), on the basis of user-individual item rankings, on the basis of item comparisons, but also in an implicit fashion, for example, by observing user interaction patterns when interacting with the recommender system \cite{GARCIA2012155}. In the following, the elicited user preferences are ranked, for example, with aggregation functions (e.g., least misery or average). Such a way of preference elicitation can work well in standard settings but can be suboptimal in informal group decision settings where users prefer to articulate their preferences and feedback in an informal fashion, i.e., not in terms of ratings. More advanced elicitation approaches leverage conversational interfaces to capture individual inputs with higher engagement, for example, user preferences can be collected in terms of a \emph{pro/con analysis} which is then used as an input for the determination of a group recommendation \cite{Felfernigetal2018EventHelpr}. However, challenges still remain in mediating conflicts and interpreting potentially ambiguous inputs.

\textbf{LLM-related Potentials.} LLMs offer the potential of providing a more natural and context-rich way of \emph{interactive preference elicitation}. On the basis of LLMs, basic item ratings can be replaced by \emph{free-form conversation} with the possibility of extracting \emph{more nuanced user preferences} \cite{FengetalLLMGroupRecInfSci2025}. For example, for informally articulated user preferences regarding a movie to watch, an LLM prompt “What kind of movie would everyone enjoy tonight?” could provide a sentiment estimate, detect conflicts, and propose \emph{explanations in real-time}. Beyond existing group recommendation approaches, LLMs can help to detect disengaged users and try to involve them in the group decision process. Furthermore, LLMs can support \emph{conflict resolution}, for example, by engaging users into further discussions with the goal to find a compromise. Such preference elicitation interfaces can be \emph{multi-modal}, i.e., combining text-based preferences with information from video streams and audio tracks from the current session \cite{LubosetalLLMMeetingAnalysis2025UMAP}.

\section{Explanations in Group Recommender Systems}

Major objectives of providing explanations in recommender systems are to improve \emph{transparency} (i.e., to help users understand the system output in terms of \emph{why?} explanations) and create \emph{trust} \cite{TintarevMasthoffExplanations2007}. In group recommender systems, explanations play an even more important role, since explanations need to help mediate between diverging user preferences and justify tradeoffs made during the phase of preference aggregation. Explanations in the group recommendation context can be based on \emph{scores} (e.g., this restaurant is recommended since it has the highest average rating). Following this approach, explanations are easy to understand. However, major properties such as conflicts, group dynamics, and compromises are not taken into account. Following a \emph{feature-based} explanation approach, features are used to explain a specific recommendation. A related example is:  ``\textit{This movie is recommended since it satisfies at least two members’ preferences for comedy and one member’s preference for history.}'' 

An important aspect in explanation generation is to present explanations \emph{without revealing sensitive user preferences}. Obfuscating preferences can lead to lower clarity of recommendations, but support privacy. The contents of Table~\ref{tab:explanation} can be used for explanation purposes by offering an aggregated view on the preferences of the group and thus respecting user boundaries by abstracting preference information to avoid personal attribution. Such explanations are also denoted as \emph{group-level explanations}, which reflect the preference landscape of the whole group. In contrast, \emph{personalized explanations} refer to the preferences of a single group member and are often not shown to the other group members. A  challenge in explanation generation is to find a  balance between group-level and personalized explanations.

\begin{table*}[h]
\centering
\caption{Aggregated feature weights as a basis for group-level explanations.}
\label{tab:explanation}
\begin{tabular}{lccc}
\toprule
\textbf{Feature} & \textbf{Average Group Weight} & \textbf{Impact on Recommendation} \\
\midrule
Action Genre & 0.75 & High \\
Romance Genre & 0.40 & Medium \\
Comedy Genre & 0.60 & High \\
Historical Setting & 0.30 & Low \\
\bottomrule
\end{tabular}
\end{table*}

\textbf{LLM-related Potentials.} LLMs can generate context-sensitive and \emph{more naturalistic explanations} \cite{LubosetalExplanationsLLMs2024}. In group decision contexts, specific aspects of the decision process have to be taken into account, for example, the sentiments of different group members, potential compromises in conflicting situations, and social relationships between group members. Due to their generative capabilities, LLMs can \emph{adapt the granularity} of an explanation based on different user roles in a decision process. For example, a detailed explanation is provided to engaged users with a corresponding topic-wise background knowledge, whereas only a summary is provided for other users with low related expertise. Explanations can also be provided for resolved conflicts and corresponding compromises, for example, ``\textit{since two members prefer hiking and one prefers city tours, we recommend a trip that includes both activities.}''

\section{Psychological Decision Models in Group Recommender Systems}

Psychological models of human decision making allow insights on which patterns of decision making groups typically follow.  Taking into account such models can help to significantly enhance the effectiveness and quality of decision support \cite{AtasetalGroupPolarization2018,Botaetal2024,ChenetalHumanDecisionmaking2013,Contrerasetal2021}. A related example concept is denoted as \emph{emotional contagion} \cite{Masthoffemotionalcontagion2006}, which represents the idea that the mood of one group member can influence the mood of other group members. Consequently, if a specific group member is very enthusiastic about an alternative (e.g., a tourist destination to visit), this could have an impact on other group members in such a way that they increase their evaluation of the mentioned alternative. If group recommender systems have knowledge about the social relationships between different users, this information can be exploited by automatically updating the corresponding preferences. Such social relationships can be represented in terms of \emph{emotional alignment scores} \cite{ZhangetalEmotionalAlignmentRecSys2024} as shown in Table \ref{tab:emotional_convergence}. Application domains where such an emotional resonance plays an important role are entertainment (e.g., which movie to watch?) and restaurant decisions -- in such contexts, emotional resonance can  affect group satisfaction \cite{Tranetal2021HumanizedRecSys}. 

\begin{table}[h]
\centering
\caption{Emotional alignment scores among group members: there is a strong emotional alignment between  $A$ and $B$.}
\label{tab:emotional_convergence}
\begin{tabular}{lccc}
\toprule
\textbf{User} & \textbf{User A} & \textbf{User B} & \textbf{User C} \\
\midrule
User A & 1.00 & 0.85 & 0.60 \\
User B & 0.85 & 1.00 & 0.50 \\
User C & 0.60 & 0.50 & 1.00 \\
\bottomrule
\end{tabular}
\end{table}

\begin{table*}[h]
\centering
\caption{Preference indicators from multimodal video data.}\label{tab:preferenceelicitation}
\begin{tabular}{|c|c|c|}
\hline
\textbf{Modality} & \textbf{Feature} & \textbf{Preference Indicator} \\
\hline
Facial Expression & Smiling duration & Positive sentiment \\
Eye Gaze & Focus on screen/object & Attention/interest \\
Voice Tone & Pitch variation & Excitement/frustration \\
Body Posture & Leaning forward & Engagement \\
\hline
\end{tabular}
\end{table*}

Another related concept denoted as \emph{groupthink} \cite{ESSER1998116} represents situations where the desire for harmony or conformity has a negative impact on the quality of a decision outcome. In the context of group recommender systems, such situations can occur if dominant group members/users (e.g., leaders on the basis of their role in the group) implicitly override the real preferences of other users. In online recommendation, groupthink might not be detected if only basic preference data is available -- a recommender system might infer group consensus. To address such issues, a recommender needs to be aware of user interaction patterns. For example, users always follow the proposed rating of a specific other person, or users rarely give feedback (or only positive feedback) on the evaluations of other users.

\emph{Group polarization} is the effect that groups as a whole take more extreme decisions compared to their individual preferences \cite{AtasetalGroupPolarization2018}. For example, if individual group members have an interest in thriller movies, a group as a whole might choose a horror movie which is in some sense more extreme than the individual group members' preferences. A similar situation might occur in the context of credit decisions where the preparedness to take risks of a whole group exceeds the preparedness to take risks of individual group members \cite{AtasetalGroupPolarization2018}. Consequently, it is important to be able to identify such decision patterns, which can be done, for example, by analyzing preference changes over time by individual group members and the divergence between the final decision and the initially articulated preferences. With such an analysis, an offering of more moderate compromise options can help to counteract polarization.

\textbf{LLM-related Potentials.} LLMs can help to more deeply integrate psychological decision models into group recommender systems by \emph{interpreting the subtle signals embedded in natural language and dialog structure}. In this context, emotional contagion can be detected, for example, by analyzing the \emph{sentiments in group chats} of individual group members throughout the decision process. In a similar fashion, LLMs can help to infer groupthink by \emph{analyzing diversity/dissent in the opinions of group members}. To counteract groupthink, an LLM can generate alternative options (not considered up to now) and also evaluate existing options differently. Given a chat history or other forms of interaction patterns (e.g., a real-time video sequence of a group decision session \cite{LubosetalLLMMeetingAnalysis2025UMAP}), group members can be analyzed with regard to their role in the decision process, i.e.,  who is the moderator, who provides ideas, who initiates detailed discussions, and who remains silent in most of the cases. Such information can be exploited for behavioral modeling and determining countermeasures that help to avoid decision biases \cite{LubosetalLLMMeetingAnalysis2025UMAP}.

\section{Research Issues in LLM Integration}

The integration of LLMs into group recommender systems triggers a couple of research issues. 

\paragraph{Recommendation of Decision Strategies.} Existing preference aggregation strategies are not appropriate in every case and there does not exist a general set of rules that clearly specify when to apply which strategy \cite{Jamesonetal2022}. LLMs can help out  by analyzing the decision context and the preferences/roles of the individual group members. With this information, LLMs can propose corresponding decision strategies. A related research challenge is the automated design, validation, and explanation of proposed decision strategies (also for  increasing trust in strategy recommendations). 

\paragraph{Assuring Fairness in LLM-based Recommendations.} On the basis of diverse and potentially contradicting preferences of individual group members, LLMs might have a tendency to over-represent the more dominant opinions in a group decision setting. To counteract such situations (specifically in the context of decision scenarios with culturally diverse groups), bias mitigation mechanisms have to be integrated into LLM architectures or corresponding post-processing components \cite{Tommasel2024FairnessGroupRecLLM}.

\paragraph{Supporting Realtime Group Modeling.} LLMs offer the capability of so-called adaptive group modeling, where new statements of individual group members can lead to a new interpretation of the overall status of a group decision process. For example, real-time video-based group decision making \cite{LubosetalLLMMeetingAnalysis2025UMAP} can profit from LLM capabilities to analyze the current status of the decision process by immediately providing updates on the basis of the interactions of group members. If one group member proposes a new decision alternative and provides an argument for or against a specific alternative, an LLM-based group recommender could immediately react in terms of indicating similar alternatives proposed in the past with arguments that led to rejection.

\paragraph{Multi-modal Preference Elicitation.} Many group recommender systems are based on the idea of collecting preferences on the basis of simple text-based (e.g., critiquing) or numeric user feedback \cite{felfernig2024group}. However, in group decision scenarios, there often exist different sources for the identification of user preferences \cite{Alvaradoetal2022}. Preferences can be inferred from direct user feedback, as well as from visual, audio, and behavioral data (e.g., user navigation behavior). For example, take a look at the example depicted in Table \ref{tab:preferenceelicitation} that focuses on preference elicitation from video streams. These basic indicators of user behavior can be interpreted by an LLM for the purpose of identifying the next relevant tasks to be performed in the group decision process.

\paragraph{Ethical and Privacy Issues.} Preference extraction from multiple data sources, specifically from video feeds and personal chats, requires the inclusion of privacy-preserving mechanisms that proactively help to avoid an unintended "transfer" of sensitive data \cite{GeetalTrustworthyRecsys2024}. Furthermore, fine-grained user consent models and intuitive preference management interfaces need to be provided for group decision scenarios.

\section{Conclusions} 
Group recommender systems provide decision support support features in  domains such as entertainment and software engineering. However, the application of group recommenders is still underrepresented in commercial contexts for various reasons, ranging from limited technological support, the need for sharing potentially sensitive data, and the danger of manipulated decision outcomes triggered by hidden agendas. This paper includes a short overview of the state of the art in group recommendation with a specific focus on the potential of integrating LLMs into group recommender system-related algorithms, preference elicitation mechanisms, explanations, and psychological models.

\section{Acknowledgements}

The presented work has been developed within the research project \textsc{GenRE} (Generative AI for Requirements Engineering) funded by the Austrian Research Promotion Agency (project number $915086$).

\bibliography{references}

\end{document}